\documentclass[11pt, a4paper]{amsart}
\usepackage{hyperref}
\usepackage{graphicx}
\usepackage{amssymb}
\usepackage[hmargin=3cm,vmargin={3.5cm,4cm}]{geometry}
\newcommand{\be}{\begin{eqnarray}}
\newcommand{\ee}{\end{eqnarray}}
\newcommand{\R}{\mathbb{R}}
\newcommand{\C}{{\mathbb{C}}} 
\newcommand{\N}{{\mathbb{N}}}

\begin{document}
%
%
%
%
\title{Wave Dispersion in the Linearised 
 Fractional Korteweg~--~de Vries equation}

\date{}

\author{Ivano Colombaro$^1$}
		\address{${}^1$ Department of Information and Communication Technologies, Universitat Pompeu Fabra. C/Roc Boronat 138, 08018, Barcelona, SPAIN.}
		\email{ivano.colombaro@bo.infn.it}
	
	    \author{Andrea Giusti$^2$}
		\address{${}^2$ Department of Physics $\&$ Astronomy, University of 	
    	    Bologna and INFN. Via Irnerio 46, 40126, Bologna, ITALY,
    	    and
    	    Arnold Sommerfeld Center, Ludwig-Maximilians-Universit\"at, Theresienstra{\ss}e 37, 80333, M\"unchen, GERMANY }
		\email{andrea.giusti@bo.infn.it}
	
    \author{Francesco Mainardi$^3$}
    	    \address{${}^3$ Department of Physics $\&$ Astronomy, University of 	
    	    Bologna and INFN. Via Irnerio 46, 40126, Bologna, ITALY.}
			\email{francesco.mainardi@bo.infn.it}

    \keywords{Dispersion, Waves, Viscoelasticity, Velocity}

    \thanks{\textbf{PACS}: 83.60.Bc, 83.80.Hj, 83.60.Uv, 62.30.+d, 02.30.Uu. \textbf{MSC}: 74J05, 76A10, 44A05, 35Q53  
    }

    \date  {\today}

\begin{abstract}
In this paper we discuss some properties of linear fractional dispersive waves. In particular, we compare the dispersion relations emerging from the kinematic wave   equation and from the linearised Korteweg~--~de Vries  equation with the corresponding time-fractionalized versions. For this purpose, we evaluate the expressions for the phase velocity and for the group velocity, highlighting the differences not only analytically, but also by means of illuminating plots.
\end{abstract}

\maketitle

\begin{center}
This paper is based on a short communication presented at the\\
\textit{19th International Conference on  Mathematical and Computational Methods in Science and Engineering}
\textit{(MACMESE '17), Berlin, Germany,  March 31 -- April 2, 2017}
 \\
 and published as a short paper on
 \\
 \textit{WSEAS Transaction of Systems, Volume 16, 2017, pp. 43-46.}
\end{center}

%
%
\section{Introduction}
\label{sec-intro}
	Linear dispersive waves are defined as physical phenomena for which the relation that connects the wave number $k$ with the angular frequency 
	$\omega$ is non-trivial. This leads to different dependences in the behaviours of the phase velocity $v _p$ and of the group velocity $v_g$ as we vary the wave number. 
	
	In general, the relation between $\omega$ and $k$, known as the dispersion relation, takes the form
	\be \label{eq-dispersion}
	\mathcal{D} (\omega , k) = 0 \, ,
	\ee 
	where $\mathcal{D}$ is a suitable real function of $\omega$ and $k$. Such a relation is, in general, satisfied by certain $\omega, \, k \in \C$. 

	Let us assume that (\ref{eq-dispersion}) can be solved explicitly in terms of a real variable ($k$ or $\omega$) by means of complex valued branches:
	\be
	\overline{\omega} _\ell (k) &\!\! \in \!\!& \C \, , \,\,\,\, k \in \R \, ,\\
	\overline{k} _m (\omega) & \!\! \in \!\!& \C \, , \,\,\,\, \omega \in \R \, ,
	\ee	
	where $\ell , \, m$ are two positive integers called mode indices. These branches provide the so-called Normal Mode Solutions for our physical system
	\be 
	\varphi _\ell (t, x; k) &\!\!=\!\!& \hbox{Re} \left\{ A _\ell (k) \, \exp\left[ i (\overline{\omega} _\ell \, t - k \, x) \right] \right\} \, , \label{conf-1} \\ 
	\varphi _m (t, x; \omega) &\!\!=\!\!& \hbox{Re} \left\{ A _\ell (\omega) \exp\left[ i (\omega \, t - \overline{k} _m \, x) \right] \right\} \, . \label{conf-2}
	\ee
	For sake of simplicity, in the following we will denote a normal mode simply by $\varphi _\ell (k)$ and $\varphi _m (\omega)$ so dropping the dependence on the space-time coordinates $x, \, t$, respectively.
	
	The normal mode solutions represent a sort of pseudo-monochromatic modes since generally they are not sinusoidal in both space and time.
	
	Now, for sake of brevity, we will omit the mode labels. Then we define, for the two cases (\ref{conf-1}) and (\ref{conf-2}) respectively, the phase velocity as
	\be 
	v_p (k) := \frac{\hbox{Re}\, \overline{\omega} (k)}{k} \, , \label{eq-vp-gen} \\
	v_p (\omega) := \frac{\omega}{\hbox{Re}\, \overline{k} (\omega)} \, .
	\ee
	In this paper, we will consider the relation (\ref{eq-vp-gen}) for the phase velocity, depending on $k$.
	
	Furthermore, we define for both cases the group velocity as
	\be 
	v_g (k) :=  
	\frac{\partial}{\partial k} \, \hbox{Re}\,\overline{\omega} (k) \, .
	\ee	
	
	Despite the fact that the theory of linear dispersive waves is a very well established and developed field of mathematical physics, the effects of fractional extensions of such linear systems on the dispersion of waves can still represent an interesting, and utterly non-trivial, research topic. The aim of this paper is to present some examples of dispersion relations related to fractional properties of mechanical systems.
	
	Particularly, in Section \ref{sec-wave} we introduce the problem of dispersion for the simple case of the kinematic wave equation. Then, in Section \ref{sec-KdV} we deal with waves satisfying the linearised Korteweg~--~de Vries (KdV) equation.

\section{The kinematic wave equation}
\label{sec-wave}

We first introduce the problem of dispersion in fractional viscoelasticity showing the case of the kinematic wave equation.

The well-known kinematic wave equation usually found in literature is

\be \label{eq:kin-ord-we}
\frac{\partial u (x,t)}{\partial t} + c_0 \frac{\partial u (x,t)}{\partial x} =0 \,,
\ee
where the velocity of the waves $c_0$ is set to one in the following for sake of simplicity.

This equation leads to a dispersion equation
\be
\omega = k \,,
\ee
from which one can easily infer $v_p=v_g$. Therefore, this is a clear example of a non-dispersive scenario.

We can appreciate a different behavior replacing the time derivative with the fractional derivative of order $\alpha$.
Applying this change, our wave equation, will take the form

\be
D^\alpha_t u (x,t) + \frac{\partial u (x,t)}{\partial x} =0 \,,
\ee
where $D^\alpha_t$ is the well known fractional Caputo derivative of order $\alpha$ (see \cite{Mainardi-book}), defined, for a certain function of time $f(t)$,
\be
D^\alpha _t f(t) = \frac{1}{\Gamma (n-\alpha)} \int _{-\infty}^t \frac{f^{(n)} (\tau)}{(t-\tau)^{\alpha +1 -n}} \, d\tau \, ,
\ee
where, in general $n \in \N $ such that 
$n-1 < \alpha <n$. In this case, we consider values $0<\alpha<1$, so $n=1$.
Thus, we can write \eqref{eq:kin-ord-we} in the Fourier domain by means of the relations
\be
D^\alpha _t & \div \quad (i \omega)^{\alpha} \, , \\
\frac{\partial }{\partial x} & \div \quad (-ik) \, ,
\ee

and the dispersion relation becomes
\be
	\overline{\omega} (k)= i^{-1+1/\alpha} k^{1/\alpha} \,.
\ee

Thus, the angular frequency presents both a real and an imaginary part. Indeed, for $k>0$, 
\be
\hbox{Re} \, \overline{\omega} (k) &=  \cos \left( \left( \frac{1}{\alpha} -1 \right) \frac{\pi}{2} \right) k^{1/\alpha} \,, \\
\hbox{Im} \, \overline{\omega} (k) &=  \sin \left( \left( \frac{1}{\alpha} -1 \right)\frac{\pi}{2} \right) k^{1/\alpha} \,.
\ee

At this point, we can easily evaluate the velocities, respectively the \textit{complex} phase velocity
\be
\label{eq-wave-vp}
\overline{v_p} (k)= i^{-1+1/\alpha} k^{-1+1/\alpha } \, ,
\ee
and \textit{complex} the group velocity
\be
\label{eq-wave-vg}
\overline{v_g} (k) = \frac{i^{-1+1/\alpha}}{\alpha} k^{-1+1/\alpha } \,.
\ee

It is then important to remark that one can immediately infer that a value of $\alpha \ne 1$ introduces dispersion effects.

\subsection{Numerical Results}

Comparing the plots of certain relevant quantities can therefore be useful to understand the phenomenon. Firstly, it could be helpful to separate the real value and the imaginary value of the expressions (\ref{eq-wave-vp}) and (\ref{eq-wave-vg}).
Indeed, one immediately finds that

\be \label{eq:kw_real_vp}
v_p (k) = \hbox{Re} \, \overline{v_p} (k) = \cos \left( \left( \frac{1}{\alpha} -1 \right) \frac{\pi}{2} \right) k^{-1 +1/\alpha} \,, \\
\hbox{Im} \, \overline{v_p} (k) = \sin \left( \left( \frac{1}{\alpha} -1 \right) \frac{\pi}{2} \right) k^{-1 +1/\alpha} \,,
\ee
 and
\be \label{eq:kw_real_vg}
v_g =\hbox{Re} \, \overline{v_g} (k) = \frac{1}{\alpha}\cos \left( \left( \frac{1}{\alpha} -1 \right) \frac{\pi}{2} \right) k^{-1 +1/\alpha} \,, \\
\hbox{Im} \, \overline{v_g} (k) = \frac{1}{\alpha}\sin \left( \left( \frac{1}{\alpha} -1 \right) \frac{\pi}{2} \right) k^{-1 +1/\alpha} \,.
\ee

\begin{figure}[htb]
\centering
\includegraphics[width=9cm]{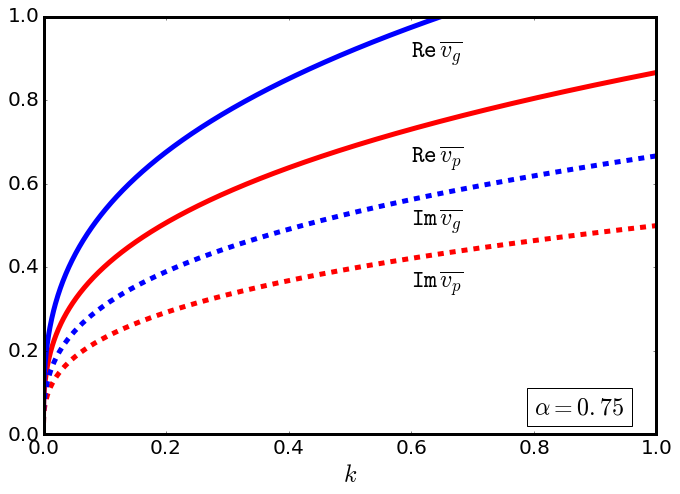}
\caption{{\small Comparison between phase velocity and group velocity, for the kinematic wave equation with fractional derivative of order $\alpha=0.75$. The straight lines represent real values, the dashed lines represent imaginary values.}}
\label{wave-0.75}
\end{figure}

\begin{figure}[htb]
\centering
\includegraphics[width=9cm]{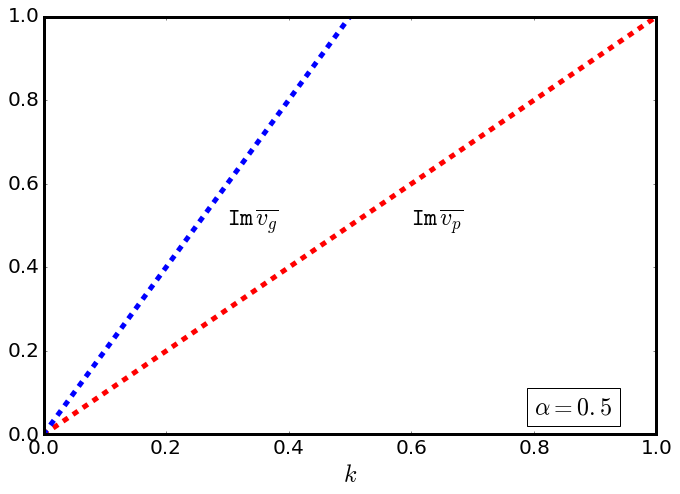}
\caption{{\small Comparison between phase velocity and group velocity, for the wave equation with fractional derivative of order $\alpha=0.5$. For $\alpha = 0.5$ the two velocities are purely imaginary functions of the wave number, so the wave vanishes.}}
\label{wave-0.5}
\end{figure}

From \figurename~\ref{wave-0.75} and \figurename~\ref{wave-0.5} we can qualitatively estimate the differences respectevely for $\alpha=0.75$ and $\alpha=0.5$. Interestingly, one finds that the real part vanishes for certain values of $\displaystyle \alpha = 1 / 2(m+1) $, where $m$ is an integer number (e.g. $\alpha=1/4$, $\alpha=1/2$).

In fact, while in \figurename~\ref{wave-0.75} we can appreciate the difference between real and imaginary part of both phase and group velocities, in \figurename~\ref{wave-0.5} is an example when the wave disappears.

\newpage

\section{The Korteweg~--~de Vries equation}
\label{sec-KdV} 

Now, we discuss a similar situation for the KdV equation. It is a non-linear equation with several applications, such as in the study of waves on shallow water surfaces (see \cite{Debnath}) or solitons descriptions (see \cite{Zabusky}). The most general form of KdV equation is

\be
\frac{\partial u (x,t)}{\partial t} + c_0 \frac{\partial u (x,t)}{\partial x} + \lambda u (x,t) \frac{\partial u (x,t)}{\partial x}+\mu \frac{\partial ^3 u (x,t)}{\partial x^3}=0
\ee
where $\lambda$, $\mu$ and $c_0$ are real numbers, as argued in \cite{Debnath-Bhatta}.

However, in this paper, we will deal the linearised KdV equation, that we recover setting $\lambda = 0$, namely

\be \label{eq-linKdV}
\frac{\partial u (x,t)}{\partial t} + \frac{\partial u (x,t)}{\partial x} + \frac{\partial ^3 u (x,t)}{\partial x^3}=0 \,,
\ee
fixing also $c_0 = \mu = 1$ for sake of simplicity. In this way the linearised KdV equation is presented as the  kinematic wave equation, with a dispersive perturbation term of the third order in space.

We can now focus our attention on the waves described by the related dispersion relation

\be
\omega (k) = k - k^3 \,.
\ee

Thanks to the latter equation it is not difficult to compute the phase velocity
\be
v_p (k) = 1 -k^2
\ee
and the group velocity
\be
v_g (k) = 1 -3k^2 \,.
\ee

It is worth remarking that, in this case, we have dispersive effects even for the unmodified wave equation.

Now, following a procedure akin to the one discussed in the previous section, we get

\be
D_t^\alpha u (x,t) + \frac{\partial u (x,t)}{\partial x} + \frac{\partial ^3 u (x,t)}{\partial x^3}=0 \,.
\ee

The resulting dispersion relation reads
\be
\overline{\omega} (k) = i^{-1+1/\alpha} \, \left( k - k^3\right)^{1/\alpha} \,,
\ee
and, again, the angular frequency can virtually be a complex number with non-vanishing imaginary part. 

Once more, from this expression of ${\omega} (k)$ we get, 
\be
\overline{v_p} (k) = i^{-1+1/\alpha} \, \left( k^{1-\alpha} - k^{3-\alpha} \right)^{1/\alpha} \, ,
\ee
and for the group velocity
\be
\overline{v_g} (k) = \frac{i^{-1+1/\alpha}}{\alpha} \, \left( k - k^{3} \right)^{1/\alpha -1} \left(1-3k^2\right) \,,
\ee
which can be split as follows
\be \label{eq:kdv_real_vp}
v_p (k) = \hbox{Re} \, \overline{v_p} (k) = \cos \left( \left( \frac{1}{\alpha} -1 \right) \frac{\pi}{2} \right)
 \left( k^{1-\alpha} - k^{3-\alpha} \right)^{1/\alpha} \,, \\
\hbox{Im} \, \overline{v_p} (k) = \sin \left( \left(\frac{1}{\alpha} -1 \right) \frac{\pi}{2} \right)
 \left( k^{1-\alpha} - k^{3-\alpha} \right)^{1/\alpha} \,,
\ee
 and
\be \label{eq:kdv_real_vg}
v_g = \hbox{Re} \, \overline{v_g} (k) = \frac{1}{\alpha}\cos \left( \left(\frac{1}{\alpha} -1 \right) \frac{\pi}{2}\right)
 \left( k - k^{3} \right)^{-1 +1/\alpha} \left(1-3k^2\right) \,, \\
\hbox{Im} \, \overline{v_g} (k) = \frac{1}{\alpha}\sin \left( \left(\frac{1}{\alpha} -1 \right) \frac{\pi}{2} \right)
 \left( k - k^{3} \right)^{-1+1/\alpha} \left(1-3k^2\right) \,.
\ee

\subsection{Numerical Results}

\begin{figure}[htb]
\centering
\includegraphics[width=9cm]{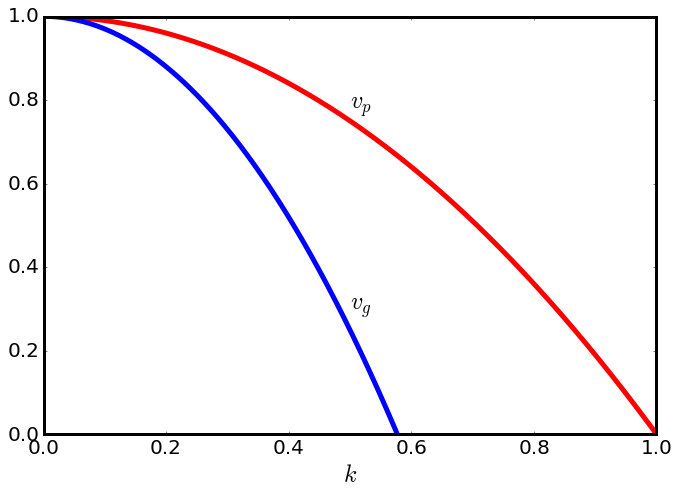}
\caption{{\small Comparison between phase velocity and group velocity, for the linearised KdV equation with ordinary derivative.}}
\label{KdV_ord}
\end{figure}

\begin{figure}[htb]
\centering
\includegraphics[width=9cm]{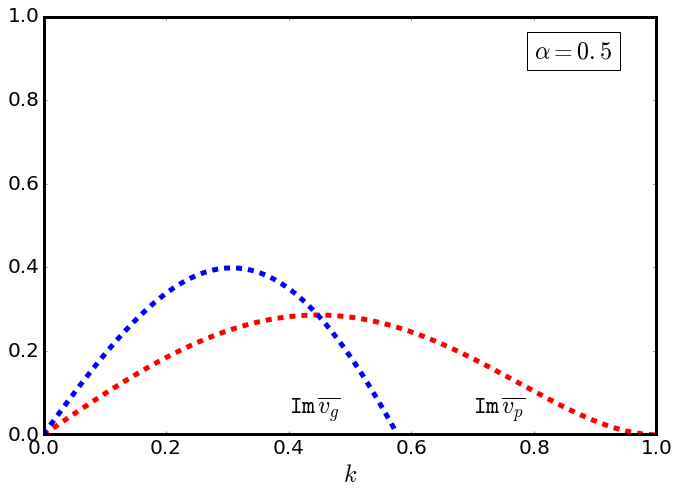}
\caption{{\small Comparison between phase velocity and group velocity, for the linearised KdV equation with frational derivative of order $1/2$, when phase and group velocities have are totallu imaginary.}}
\label{KdV_1/2}
\end{figure}

\begin{figure}[htb]
\centering
\includegraphics[width=9cm]{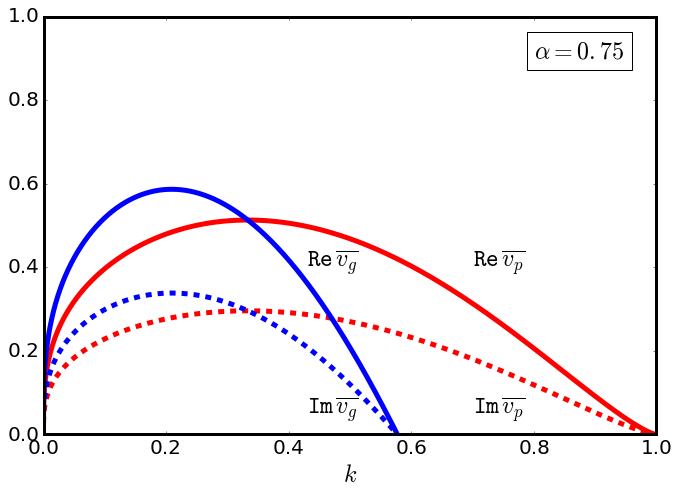}
\caption{{\small Comparison between phase velocity and group velocity, for the linearised KdV equation with fractional  derivative of order $\alpha=3/4$.}}
\label{KdV_3/4}
\end{figure}

We can conclude that for $\alpha=1$, so in the ordinary case, phase velocity and group velocity are real-valued, as well shown by the \figurename~\ref{KdV_ord}, and totally imaginary-valued for other values of $\alpha$, as can be stated looking at \figurename~\ref{KdV_1/2}.
In particular, it is remarkable that the argument of the cosine functions in \eqref{eq:kdv_real_vp} and \eqref{eq:kdv_real_vg} is equivalent to the corresponding one for the kinematic wave, emerging in \eqref{eq:kw_real_vp} and \eqref{eq:kw_real_vg}.
So, for $\displaystyle \alpha = 1/ 2(m+1) $, with $m \in \N$, velocities do not present real part.

For other real values of $\alpha$, we find a mixed behavior, as we can see from \figurename~\ref{KdV_3/4}.


Furthermore, we can note that there is a point where $\hbox{Re}\, \overline{v_p} (k) = \hbox{Re}\, v_g (k)$ and a point where $\hbox{Im}\, \overline{v_p} (k) = \hbox{Im}\, v_g (k)$.

\newpage

\section{Conclusion}
\label{conclusion}

	In conclusion, it seems that the procedure of fractionalizing a linear wave equation leads to major modifications of the corresponding dispersion relation. 
	
	This analysis can surely be extended further by considering fractional derivative with respect to the space spatial coordinate,  however this discussion is left for future investigations.

\end{document}